# The expectation value of the field operator.

Dan Solomon
University of Illinois
Chicago, IL
dsolom2@uic.edu
June 13, 2014

## Abstract.

Much of the mathematical development of quantum field theory has been in support of determining the S-matrix in order to calculate scattering cross sections. However there is also an interest in determining how expectation values of field operators evolve in time from an initial state. In this paper I will examine some aspects of this problem.

## 1. Introduction

Much of the mathematical development of quantum field theory involves the S-matrix which is used to obtain scattering cross sections. In this case the mathematical formulation attempts to determine the probability of some initial state evolving into some final state where both the initial and final states are known. However there is also interest in determining how the expectation values associated with the quantum field evolve in time [ 1,2,3,4,5]. In this case we are only interested in the initial state of the system and, in addition, we need an equation that determines how the expectation values evolve in time from some initial state. In this paper we will address some aspects of this problem. The system under consideration will be a real scalar field.

## 2. The Heisenberg picture

We will start out in the Heisenberg picture. In this case the state vectors are time independent and all the time dependence of the system is associated with the field operators. Assume that the classical action of the system is given by,

$$S = \int dt \int d^3x \left\{ \frac{1}{2}\varphi\left(\partial_\mu \partial^\mu - m^2\right)\varphi - V(\varphi) \right\} \tag{2.1}$$

where $V(\varphi)$ is a polynomial expansion in $\varphi(x)$ where $x$ stands for the quantity $(\vec{x},t)$ in 3-1 dimensions. Also natural units are assumed so that $\hbar = c = 1$. At times we will consider a "model system" which will be $\varphi^4$ theory in the presence of an external source $s(x)$. For this case,

$$V(\varphi) = (\lambda/4!)\varphi^4 + s\varphi. \tag{2.2}$$

In the Heisenberg picture the field operators are $\hat{\varphi}_H(x)$ and $\hat{\pi}_H(x)$. The field operators satisfy the equal time commutation relationships,

$$\left[\hat{\varphi}_H(\vec{x}_1,t),\hat{\pi}_H(\vec{x}_2,t)\right]=i\delta^3(\vec{x}_1-\vec{x}_2);\quad \left[\hat{\varphi}_H(\vec{x}_1,t),\hat{\varphi}_H(\vec{x}_2,t)\right]=0;\quad \left[\hat{\pi}_H(\vec{x}_1,t),\hat{\pi}_H(\vec{x}_2,t)\right]=0 \quad (2.3)$$

They obey the following equations which correspond to the classical equations of motion,

$$D_x\hat{\varphi}_H-V'(\hat{\varphi}_H)=0;\quad \hat{\pi}_H=\frac{\partial\hat{\varphi}_H}{\partial t} \quad (2.4)$$

where $D_x$ and $V'$ are defined by,

$$D_x=\partial_\mu\partial^\mu-m^2 \text{ and } V'(\varphi)=\frac{\partial V(\varphi)}{\partial\varphi} \quad (2.5)$$

and where $\partial_\mu\partial^\mu=-\frac{\partial^2}{\partial t^2}+\left|\vec{\nabla}\right|^2$. For example for our model system,

$$D_x\hat{\varphi}_H(x)-\frac{\lambda}{3!}\hat{\varphi}_H^3(x)=s(x) \quad (2.6)$$

Let $|\Omega\rangle$ be the state vector of the system. Assume that it is normalized, that is, $\langle\Omega|\Omega\rangle=1$. Since we are working in the Heisenberg picture the state vector is time independent. Let us suppose that we are interested in the expectation value of the field operator which is $\langle\Omega|\hat{\varphi}_H(x)|\Omega\rangle$. Multiply Eq. (2.6) from the left by $\langle\Omega|$ and from the right by $|\Omega\rangle$. From this we obtain,

$$D_x\langle\Omega|\hat{\varphi}_H(x)|\Omega\rangle-\frac{\lambda}{3!}\langle\Omega|\hat{\varphi}_H^3(x)|\Omega\rangle=s(x) \quad (2.7)$$

Consider the case where $\lambda=0$. In this case $D_x\langle\Omega|\hat{\varphi}_H(x)|\Omega\rangle=s(x)$. This is readily solvable. The solution is $\langle\Omega|\hat{\varphi}_H(x)|\Omega\rangle=\chi(x)+\int G(x,x')s(x')dx'$ where $G(x,x')$ is the Greens function associated with the operator $D_x$ and $\chi(x)$ satisfies the equation $D_x\chi(x)=0$. However for the case $\lambda\neq 0$ the situation is more complicated. That is because there is one equation but two "superficially" independent quantities which are $\langle\Omega|\hat{\varphi}_H(x)|\Omega\rangle$ and $\langle\Omega|\hat{\varphi}_H^3(x)|\Omega\rangle$. Of course they are not really independent because they are both dependent on the quantity $\hat{\varphi}_H(x)$ which is fully specified by Eq. (2.6). However is not readily apparent how to solve (2.7) in the form presented.

## 3. A differential equation for the expectation value.

A possible way to overcome the difficulty discussed above will the presented in this section. Define the following operator,

$$\hat{Z}_E[J] = T\exp\left(i\int_{-\infty}^{+\infty} J(x)\hat{\varphi}_H(x)dx\right) \tag{3.1}$$

where $T$ is the time ordering operator and $J(x)$ is a real valued function. In the above we have used the following convention,

$$\int_{-\infty}^{+\infty} f(x)dx \rightarrow \int d^3x \int_{-\infty}^{+\infty} f(\vec{x},t)dt \tag{3.2}$$

where integration over all of 3-dimensional space is implied by the first integral on the right. It can be shown that,

$$\frac{1}{i}\frac{\delta}{\delta J(x_1)}\frac{1}{i}\frac{\delta}{\delta J(x_2)}\cdots\frac{1}{i}\frac{\delta}{\delta J(x_n)}\mathrm{T}\exp\left(i\int_{-\infty}^{+\infty} dxJ\hat{\varphi}_H\right) = T\left\{\hat{\varphi}_H(x_1)\hat{\varphi}_H(x_2)\ldots\hat{\varphi}_H(x_n)\exp\left(i\int_{-\infty}^{+\infty} dxJ\hat{\varphi}_H\right)\right\} \tag{3.3}$$

Using this relationship and the commutations (2.3) it is shown in the Appendix that,

$$\frac{\partial^2}{\partial t^2}\frac{\delta\hat{Z}_E[J]}{i\delta J(x)} = T\left\{\left(\frac{\partial^2\hat{\varphi}_H(x)}{\partial t^2}\exp\left(i\int_{-\infty}^{+\infty} J(x)\hat{\varphi}_H(x)dx\right)\right)\right\} - J(x)\hat{Z}_E[J] \tag{3.4}$$

Referring again to (3.3) we obtain the following two relationships,

$$\frac{\partial^2}{\partial x^2}\frac{\delta}{i\delta J(x)}\hat{Z}_E[J] = T\left\{\frac{\partial^2\hat{\varphi}_H(x)}{\partial x^2}\exp\left(i\int_{-\infty}^{+\infty} J(x)\hat{\varphi}_H(x)dx\right)\right\} \tag{3.5}$$

$$V'\left(\frac{\delta}{i\delta J(x)}\right)\hat{Z}_E[J] = T\left\{V'(\hat{\varphi}_H(x))\exp\left(i\int_{-\infty}^{+\infty} J(x)\hat{\varphi}_H(x)dx\right)\right\} \tag{3.6}$$

From all this we obtain,

$$\begin{aligned}&\left\{D_x\left(\frac{\delta}{i\delta J(x)}\right) - V'\left(\frac{\delta}{i\delta J(x)}\right)\right\}\hat{Z}_E[J] = \\ &= T\left\{\left(D_x\hat{\varphi}_H(x) - V'(\hat{\varphi}_H(x))\right)\exp\left(i\int_{-\infty}^{+\infty} J(x)\hat{\varphi}_H(x)dx\right)\right\} + J(x)\hat{Z}_E[J]\end{aligned} \tag{3.7}$$

Use (2.4) in the above to obtain,

$$\left\{ D_x\left(\frac{\hbar}{i}\frac{\delta}{\delta J(x)}\right) - V'\left(\frac{\hbar}{i}\frac{\delta}{\delta J(x)}\right)\right\}\hat{Z}_E[J] = J(x)\hat{Z}_E[J] \tag{3.8}$$

Next assume we have two state vectors $|\Omega_1\rangle$ and $|\Omega_2\rangle$. Referring to (3.8) the quantity $\langle\Omega_1|\hat{Z}_E[J]|\Omega_2\rangle$ satisfies,

$$\left\{ D_x\left(\frac{\delta}{i\delta J(x)}\right) - V'\left(\frac{\delta}{i\delta J(x)}\right)\right\}\langle\Omega_1|\hat{Z}_E[J]|\Omega_2\rangle = J(x)\langle\Omega_1|\hat{Z}_E[J]|\Omega_2\rangle \tag{3.9}$$

Next define the quantity $W[J]$ by,

$$\exp(iW[J]) = \langle\Omega_1|\hat{Z}_E[J]|\Omega_2\rangle \tag{3.10}$$

From this we obtain,

$$\exp(-iW[J])\frac{\delta}{i\delta J(x)}\exp(iW[J]) = \phi(x;J) + \frac{\delta}{i\delta J(x)} \tag{3.11}$$

where $\phi(x;J)$ is defined by,

$$\phi(x;J) = \frac{\delta W[J]}{\delta J(x)} = \frac{\langle\Omega_1|T\hat{\varphi}_H(x)\exp\left(i\int_{-\infty}^{+\infty} J(x)\hat{\varphi}_H(x)dx\right)|\Omega_2\rangle}{\langle\Omega_1|Z_E[J]|\Omega_2\rangle} \tag{3.12}$$

Note that $\phi(x;J)$ is dependent on the space-time point $x$ and the function $J(x)$. It is also dependent on the state vectors $|\Omega_1\rangle$ and $|\Omega_2\rangle$ as is $W[J]$.

Use the above results in (3.9) to obtain,

$$\left\{ D_x\phi(x;J) - V'\left(\phi(x;J) + \frac{\delta}{i\delta J(x)}\right)\right\} = J(x) \tag{3.13}$$

where we have used,

$$\frac{\delta}{i\delta J(x)}\cdot 1 = 0 \tag{3.14}$$

Similar results have been obtained elsewhere using somewhat different approaches [1,6,7,8].

For our model system $V'(\varphi)=\left[(\lambda/3!)\varphi^3(x)+s(x)\right]$ (see Eq. (2.2)). In this case,

$$\begin{aligned} V'\left(\phi(x;J)+\frac{\delta}{i\delta J(x)}\right)&=\frac{\lambda}{3!}\left(\phi(x;J)+\frac{\delta}{i\delta J(x)}\right)^3+s(x) \\ &=\frac{\lambda}{3!}\left(\phi^3(x;J)+3\phi(x;J)\frac{\delta\phi(x;J)}{i\delta J(x)}+\left(\frac{1}{i}\right)^2\frac{\delta^2\phi(x;J)}{\delta J(x)^2}\right)+s(x) \end{aligned} \tag{3.15}$$

Use this in (3.13) to obtain,

$$\left\{D_x\phi(x;J)-\frac{\lambda}{3!}\left(\phi^3(x;J)+3\phi(x;J)\frac{\delta\phi(x;J)}{i\delta J(x)}+\left(\frac{1}{i}\right)^2\frac{\delta^2\phi(x;J)}{\delta J(x)^2}\right)\right\}=s(x)+J(x) \tag{3.16}$$

Now rearrange terms to obtain,

$$\left\{D_x\phi(x;J)-\frac{\lambda}{3!}\phi^3(x;J)-C(x;J)\right\}=s(x)+J(x) \tag{3.17}$$

where,

$$C(x;J)=\frac{\lambda}{3!}\left(3\phi(x;J)\frac{\delta\phi(x;J)}{i\delta J(x)}+\left(\frac{1}{i}\right)^2\frac{\delta^2\phi(x;J)}{\delta J(x)^2}\right) \tag{3.18}$$

Note that if the term $C(x;J)$ was removed from (3.17) then this equation would be equivalent to the classical equations of motion in the presence of a source $s(x)+J(x)$. Therefore we can think of $C(x;J)$ as a "quantum correction" to the classical equations of motion. Next we will show that (3.16) is identical to Eq. (2.6) if take the limit $J(x)=0$. From (3.12) and (3.3) we obtain,

$$\frac{\delta\phi(x_1;J)}{i\delta J(x_2)}=\frac{\langle\Omega_1|T\hat{\varphi}_H(x_1)\hat{\varphi}_H(x_2)\exp\left(i\int_{-\infty}^{+\infty}J(x)\hat{\varphi}_H(x)dx\right)|\Omega_2\rangle}{\langle\Omega_1|Z_E[J]|\Omega_2\rangle}-\phi(x_2;J)\phi(x_1;J) \tag{3.19}$$

Take the variation of this with respect to $J(x_3)$ to obtain,

$$
\begin{aligned}
\frac{\delta^2\phi(x_1;J)}{i\delta J(x_3)i\delta J(x_2)} &= \frac{\left\langle\Omega_1\middle|T\hat{\varphi}_H(x_1)\hat{\varphi}_H(x_2)\hat{\varphi}_H(x_3)\exp\left(i\int_{-\infty}^{+\infty}J(x)\hat{\varphi}_H(x)dx\right)\middle|\Omega_2\right\rangle}{\left\langle\Omega_1\middle|Z_E[J]\middle|\Omega_2\right\rangle}; \\
&-\frac{\left\langle\Omega_1\middle|T\hat{\varphi}_H(x_1)\hat{\varphi}_H(x_2)\exp\left(i\int_{-\infty}^{+\infty}J(x)\hat{\varphi}_H(x)dx\right)\middle|\Omega_2\right\rangle}{\left\langle\Omega_1\middle|Z_E[J]\middle|\Omega_2\right\rangle}\phi(x_3;J) \\
&-\left(\frac{\delta\phi(x_2;J)}{i\delta J(x_3)}\phi(x_1;J)+\frac{\delta\phi(x_1;J)}{i\delta J(x_3)}\phi(x_2;J)\right)
\end{aligned}
\tag{3.20}
$$

In the limit that $J(x)=0$ and $x_1=x_2=x_3=x$ we have the following,

$$
\phi_0(x)\equiv\phi(x;J)\big|_{J=0}=\frac{\left\langle\Omega_1\middle|\hat{\varphi}_H(x)\middle|\Omega_2\right\rangle}{\left\langle\Omega_1\middle|\Omega_2\right\rangle}
\tag{3.21}
$$

$$
\left.\frac{\delta\phi(x;J)}{i\delta J(x)}\right|_{J=0}=\frac{\left\langle\Omega_1\middle|\hat{\varphi}_H(x)\hat{\varphi}_H(x)\middle|\Omega_2\right\rangle}{\left\langle\Omega_1\middle|\Omega_2\right\rangle}-\phi_0(x)\phi_0(x)
\tag{3.22}
$$

$$
\begin{aligned}
\left.\frac{\delta^2\phi(x;J)}{i\delta J(x)i\delta J(x)}\right|_{J=0} &= \frac{\left\langle\Omega_1\middle|\hat{\varphi}_H(x)\hat{\varphi}_H(x)\hat{\varphi}_H(x)\middle|\Omega_2\right\rangle}{\left\langle\Omega_1\middle|\Omega_2\right\rangle} \\
&-3\frac{\left\langle\Omega_1\middle|\hat{\varphi}_H(x)\hat{\varphi}_H(x)\middle|\Omega_2\right\rangle}{\left\langle\Omega_1\middle|\Omega_2\right\rangle}\phi_0(x)+2\phi_0(x)\phi_0(x)\phi_0(x)
\end{aligned}
\tag{3.23}
$$

Use these in (3.16) and set $J(x)=0$ to obtain,

$$
\frac{D_x\left\langle\Omega_1\middle|\hat{\varphi}_H(x)\middle|\Omega_2\right\rangle}{\left\langle\Omega_1\middle|\Omega_2\right\rangle}-\frac{\lambda}{3!}\frac{\left\langle\Omega_1\middle|\hat{\varphi}_H^3(x)\middle|\Omega_2\right\rangle}{\left\langle\Omega_1\middle|\Omega_2\right\rangle}=s(x)
\tag{3.24}
$$

Multiply both sides of the equation by $\left\langle\Omega_1\middle|\Omega_2\right\rangle$ and rearrange terms to obtain,

$$
\left\langle\Omega_1\right|\left\{D_x\hat{\varphi}_H(x)-\frac{\lambda}{3!}\hat{\varphi}_H^3(x)-s(x)\right\}\left|\Omega_2\right\rangle=0
\tag{3.25}
$$

This is equivalent to the evolution equation for the field operator given by (2.6).

Next let $\left|\Omega_2\right\rangle=\left|\Omega_1\right\rangle=\left|\Omega\right\rangle$ where $\left\langle\Omega\middle|\Omega\right\rangle=1$. In this case,

$$\phi(x;J) = \frac{\langle \Omega | T\hat{\varphi}_H(x) \exp\left( i \int_{-\infty}^{+\infty} J(x)\hat{\varphi}_H(x)\,dx \right) | \Omega \rangle}{\langle \Omega | Z_E[J] | \Omega \rangle} \tag{3.26}$$

If we set $J(x)=0$ this becomes $\phi_0(x) = \phi(x;J)\big|_{J=0} = \langle \Omega | \hat{\varphi}_H(x) | \Omega \rangle$ which is just the expectation value of the operator $\hat{\varphi}_H(x)$. The result of all this is that instead of having two "superficially" independent quantities $\langle \Omega | \hat{\varphi}_H(x) | \Omega \rangle$ and $\langle \Omega | \hat{\varphi}_H^3(x) | \Omega \rangle$ in the differential equation (2.7) we have one quantity $\phi(x;J)$ in the differential equation (3.16). In should be, in principle, possible to solve this differential equation for $\phi(x;J)$ and then take $J(x)=0$ to obtain the expectation value of the field operator which is the desired result. This will be explored further in a later paper.

## 4. An alternative expression for $Z_E$[J].

In this section we consider alternative forms of the expression for $\hat{Z}_E[J]$. In particular we want to write it in terms of the interaction picture field operators instead of Heisenberg picture operators. In the Heisenberg picture all of the time evolution of the system is given by the field operators. In the interaction picture a portion of the time evolution is contained in the field operators and the rest is in the state vector. The Interaction picture field operators $\hat{\varphi}_I(x)$ and $\pi_I(x)$ obey the equal time commutations given in (2.3) and satisfy,

$$D_x \hat{\varphi}_I = 0; \quad \hat{\pi}_I = \frac{\partial \hat{\varphi}_I}{\partial t} \tag{4.1}$$

The Heisenberg field operators can be expressed in terms of the Interaction picture field operators [9] according to the expression,

$$\hat{\varphi}_H(x) = \hat{U}(-\infty,t)\hat{\varphi}_I(x)\hat{U}(t,-\infty) \tag{4.2}$$

where $\hat{U}(t,t')$ is a unitary operator that satisfies the relationships,

$$\hat{U}(t_1,t_3) = \hat{U}(t_1,t_2)\hat{U}(t_2,t_3) \text{ and } \hat{U}^\dagger\hat{U} = \hat{U}\hat{U}^\dagger = 1 \tag{4.3}$$

and obeys,

$$i\frac{\partial}{\partial t}\hat{U}(t,t') = \hat{V}(t)\hat{U}(t,t') \tag{4.4}$$

where $\hat{U}(t,t)=1$ and,

$$\hat{V}(t)=\int V\left(\hat{\varphi}_I\left(\vec{x},t\right)\right)d^3x \tag{4.5}$$

The solution to this,

$$\hat{U}(t,t')=T\exp\left(-i\int_{t'}^{t}\hat{V}_I\left(t''\right)dt''\right) \tag{4.6}$$

Consider the quantity,

$$L_n=T\left(\hat{\varphi}_H\left(x_1\right)\hat{\varphi}_H\left(x_2\right)\ldots\hat{\varphi}_H\left(x_n\right)\right) \tag{4.7}$$

Assume that $t_1\geq t_2\geq,\ldots,\geq t_n$ then,

$$L_n=\hat{\varphi}_H\left(x_1\right)\hat{\varphi}_H\left(x_2\right)\ldots\hat{\varphi}_H\left(x_n\right) \tag{4.8}$$

Use (4.2) in the above to obtain,

$$L_n=\hat{U}\left(-\infty,t_1\right)\hat{\varphi}_I\left(x_1\right)\hat{U}\left(t_1,t_2\right)\hat{\varphi}_I\left(x_2\right)\hat{U}\left(t_2,t_3\right)\ldots\hat{U}\left(t_{n-1},t_n\right)\hat{\varphi}_I\left(x_n\right)\hat{U}\left(t_n,-\infty\right) \tag{4.9}$$

Use $\hat{U}\left(t_1,+\infty\right)\hat{U}\left(+\infty,t_1\right)=1$ and $\hat{U}\left(-\infty,t_1\right)\hat{U}\left(t_1,+\infty\right)=\hat{U}\left(-\infty,+\infty\right)$ in the above to obtain,

$$L_n=\hat{U}\left(-\infty,+\infty\right)\hat{U}\left(+\infty,t_1\right)\hat{\varphi}_I\left(x_1\right)\hat{U}\left(t_1,t_2\right)\hat{\varphi}_I\left(x_2\right)\hat{U}\left(t_2,t_3\right)\ldots\hat{U}\left(t_{n-1},t_n\right)\hat{\varphi}_I\left(x_n\right)\hat{U}\left(t_n,-\infty\right) \tag{4.10}$$

This can be written as,

$$L_n=\hat{U}\left(-\infty,+\infty\right)T\left\{\hat{U}\left(+\infty,-\infty\right)\hat{\varphi}_I\left(x_1\right)\hat{\varphi}_I\left(x_2\right)\ldots\hat{\varphi}_I\left(x_n\right)\right\} \tag{4.11}$$

Use this result in (3.1) to obtain,

$$\hat{Z}_E\left[J\right]=\hat{U}\left(-\infty,+\infty\right)T\left[\hat{U}\left(+\infty,-\infty\right)\exp\left(i\int_{-\infty}^{+\infty}J\left(x\right)\hat{\varphi}_I\left(x\right)dx\right)\right] \tag{4.12}$$

The result is that,

$$\hat{Z}_E\left[J\right]=\hat{U}\left(-\infty,+\infty\right)\hat{U}\left(+\infty,-\infty;J\right) \tag{4.13}$$

where,

$$\hat{U}\left(+\infty,-\infty;J\right)=T\left[\exp\left(-i\int_{-\infty}^{+\infty}V\left(\hat{\varphi}_I\left(x\right)\right)dx\right)\exp\left(i\int_{-\infty}^{+\infty}J\left(x\right)\hat{\varphi}_I\left(x\right)dx\right)\right] \tag{4.14}$$

From [10] this can also be expressed as,

$$\hat{U}\left(+\infty,-\infty;J\right)=T\left[\exp\left(-i\int_{-\infty}^{+\infty}dx\left[V\left(\hat{\varphi}_I\left(x\right)\right)-J\left(x\right)\hat{\varphi}_I\left(x\right)\right]\right)\right] \tag{4.15}$$

Consider the action of the operator $\hat{Z}_E\left[J\right]$ on some state vector $\left|\Omega\right\rangle$. First the operator $\hat{U}\left(+\infty,-\infty;J\right)$ evolves the system in the presence of the external source $J\left(x\right)$ from $t=-\infty$ to $t=+\infty$. Next $\hat{U}\left(-\infty,+\infty\right)$ evolves the system from $t=+\infty$ back to $t=-\infty$ with $J\left(x\right)=0$. This is similar to the in-in formulism [4,5].

From results in Ref. [10] we can write $\hat{U}\left(+\infty,-\infty;J\right)$ as expressed by (4.14) as,

$$\hat{U}\left(+\infty,-\infty;J\right)=\exp\left(-i\int_{-\infty}^{+\infty}dxV\left(\frac{1}{i}\frac{\delta}{\delta J\left(x\right)}\right)\right)T\exp\left(i\int_{-\infty}^{+\infty}J\left(x\right)\hat{\varphi}_I\left(x\right)dx\right) \tag{4.16}$$

It is also shown in Ref. [10] that,

$$T\exp\left(i\int_{-\infty}^{+\infty}J\left(x\right)\hat{\varphi}_I\left(x\right)dx\right)=:\exp\left(i\int_{-\infty}^{+\infty}J\left(x\right)\hat{\varphi}_I\left(x\right)dx\right):\exp\left(-\frac{i}{2}\int_{-\infty}^{+\infty}\int_{-\infty}^{+\infty}J\left(x\right)\Delta_F\left(x-x'\right)J\left(x'\right)dxdx'\right) \tag{4.17}$$

where the colons : : indicate that the quantity between the colons is to be put in normal order and $\Delta_F\left(x-x'\right)$ is the Feynman propagator and satisfies $D_x\Delta_F\left(x-x'\right)=\delta^4\left(x-x'\right)$. Using this result we obtain,

$$\hat{U}\left(+\infty,-\infty;J\right)=\exp\left(-i\int_{-\infty}^{+\infty}dxV\left(\frac{1}{i}\frac{\delta}{\delta J\left(x\right)}\right)\right)\left[\begin{array}{l}:\exp\left(i\int_{-\infty}^{+\infty}J\left(x\right)\hat{\varphi}_I\left(x\right)dx\right):\\ \times\exp\left(-\frac{i}{2}\int_{-\infty}^{+\infty}\int_{-\infty}^{+\infty}J\left(x\right)\Delta_F\left(x-x'\right)J\left(x'\right)dxdx'\right)\end{array}\right] \tag{4.18}$$

We can use this result to evaluate $\hat{U}\left(-\infty,+\infty\right)$ which becomes,

$$\hat{U}\left(-\infty,+\infty\right)=\hat{U}^{\dagger}\left(+\infty,-\infty;J\right)\Big|_{j=0} \tag{4.19}$$

## 5. Conclusion.

To summarize, in Eq. (3.1) we defined the operator $\hat{Z}_E\left[J\right]$. It was then shown that this operator satisfied the differential equation (3.8). Using this result it was shown in the rest of Section 3 how to derive the differential equation (see (3.13) or (3.16)) for the quantity $\phi(x;J)$. This differential equation consists of a part that satisfies the classical equations of motion to which is added a quantum correction. In the limit that $J(x)=0$ the quantity $\phi(x;J)$ will be equal to the expectation value of the field operator $\hat{\varphi}_H(x)$. Next in Section 4 we examined some further properties of the operator $\hat{Z}_E[J]$. It was shown that the effect of acting on a state vector $|\Omega\rangle$ with the operator $\hat{Z}_E[J]$ is to evolve the state vector forward in time from $-\infty$ to $+\infty$ and in the presence of the source $J(x)$ and then to evolve backward in time from $+\infty$ to $-\infty$ during which $J(x)=0$.

# Appendix

We want to show that (3.4) is true. From (3.1),

$$\hat{Z}_E[J]=1+\sum_{n=1}^{\infty} i^n S_n \tag{5.1}$$

where,

$$S_n=\frac{1}{n!}T\int_{-\infty}^{+\infty}dx_1\int_{-\infty}^{+\infty}dx_2\ldots\int_{-\infty}^{+\infty}dx_n J(x_1)\hat{\varphi}_H(x_1)J(x_2)\hat{\varphi}_H(x_2)\ldots J(x_n)\hat{\varphi}_H(x_n) \tag{5.2}$$

This can be written as,

$$S_n=\int_{-\infty}^{+\infty}dx_1 J(x_1)\hat{\varphi}_H(x_1)\int_{-\infty}^{+t_1}dx_2 J(x_2)\hat{\varphi}_H(x_2)\ldots\int_{-\infty}^{+t_{n-1}}dx_n J(x_n)\hat{\varphi}_H(x_n) \tag{5.3}$$

Recall

$$\int_{-\infty}^{+t_n}dx f(x)\equiv\int d^3x\int_{-\infty}^{+t_n}dt f(\vec{x},t) \tag{5.4}$$

It can be shown that

$$\begin{aligned}\frac{\partial^2}{\partial t^2}\frac{\delta S_n}{\delta J(x)}=&\frac{1}{(n-1)!}T\left[\frac{\partial^2\hat{\varphi}_H(x)}{\partial t^2}\int_{-\infty}^{+\infty}dx_1\int_{-\infty}^{+\infty}dx_2\ldots\int_{-\infty}^{+\infty}dx_{n-1}J(x_1)\hat{\varphi}_H(x_1)J(x_2)\hat{\varphi}_H(x_2)\ldots J(x_{n-1})\hat{\varphi}_H(x_{n-1})\right]\\&+-iJ(x)\frac{1}{(n-2)!}T\left[\int_{-\infty}^{+\infty}dx_1\int_{-\infty}^{+\infty}dx_2\ldots\int_{-\infty}^{+\infty}dx_{n-2}J(x_1)\hat{\varphi}_H(x_1)J(x_2)\hat{\varphi}_H(x_2)\ldots J(x_{n-2})\hat{\varphi}_H(x_{n-2})\right]\end{aligned} \tag{5.5}$$

When this is used in (5.1) then we obtain (3.4)

As an example show that (5.5) hold for $S_4$ .

$$S_4 = \int_{-\infty}^{+\infty} dx_1 J(x_1)\hat{\varphi}_H(x_1) \int_{-\infty}^{+t_1} dx_2 J(x_2)\hat{\varphi}_H(x_2) \int_{-\infty}^{+t_2} dx_3 J(x_3)\hat{\varphi}_H(x_3) \int_{-\infty}^{+t_3} dx_4 J(x_4)\hat{\varphi}_H(x_4) \tag{5.6}$$

From this expression we obtain,

$$\begin{aligned}
\frac{\delta S_4}{\delta J(x)} &= \hat{\varphi}_H(x) \int_{-\infty}^{+t} dx_2 J(x_2)\hat{\varphi}_H(x_2) \int_{-\infty}^{+t_2} dx_3 J(x_3)\hat{\varphi}_H(x_3) \int_{-\infty}^{+t_3} dx_4 J(x_4)\hat{\varphi}_H(x_4) \\
&+ \int_{t}^{+\infty} dx_1 J(x_1)\hat{\varphi}_H(x_1)\hat{\varphi}_H(x) \int_{-\infty}^{+t} dx_3 J(x_3)\hat{\varphi}_H(x_3) \int_{-\infty}^{+t_3} dx_4 J(x_4)\hat{\varphi}_H(x_4) \\
&+ \int_{t}^{+\infty} dx_1 J(x_1)\hat{\varphi}_H(x_1) \int_{t}^{+t_1} dx_2 J(x_2)\hat{\varphi}_H(x_2)\hat{\varphi}_H(x) \int_{-\infty}^{+t} dx_4 J(x_4)\hat{\varphi}_H(x_4) \\
&+ \int_{t}^{+\infty} dx_1 J(x_1)\hat{\varphi}_H(x_1) \int_{t}^{+t_1} dx_2 J(x_2)\hat{\varphi}_H(x_2) \int_{t}^{+t_2} dx_3 J(x_3)\hat{\varphi}_H(x_3)\hat{\varphi}_H(x)
\end{aligned} \tag{5.7}$$

Next take the first derivative with respect to $t$ . The time $t$ appears in the $\hat{\varphi}_H(x) = \hat{\varphi}_H(\vec{x},t)$ and in some of the limits on the integrals. However the final result of the derivatives on the limit of the integrals is to produce terms of the form $\left[\hat{\varphi}_H(\vec{x},t), \int d^3x_1 J(\vec{x}_1,t)\hat{\varphi}_H(\vec{x}_1,t)\right]$ . These terms are zero due to the fact that the equal time commutator $\left[\hat{\varphi}_H(\vec{x},t), \hat{\varphi}_H(\vec{x}_1,t)\right] = 0$ . Use this fact and $\frac{\partial \hat{\varphi}_H(x)}{\partial t} = \hat{\pi}_H(x)$ to obtain,

$$\begin{aligned}
\frac{\partial}{\partial t}\frac{\delta S_4}{\delta J(x)} &= \hat{\pi}_H(x) \int_{-\infty}^{+t} dx_2 J(x_2)\hat{\varphi}_H(x_2) \int_{-\infty}^{+t_2} dx_3 J(x_3)\hat{\varphi}_H(x_3) \int_{-\infty}^{+t_3} dx_4 J(x_4)\hat{\varphi}_H(x_4) \\
&+ \int_{t}^{+\infty} dx_1 J(x_1)\hat{\varphi}_H(x_1)\hat{\pi}_H(x) \int_{-\infty}^{+t} dx_3 J(x_3)\hat{\varphi}_H(x_3) \int_{-\infty}^{+t_3} dx_4 J(x_4)\hat{\varphi}_H(x_4) \\
&+ \int_{t}^{+\infty} dx_1 J(x_1)\hat{\varphi}_H(x_1) \int_{t}^{+t_1} dx_2 J(x_2)\hat{\varphi}_H(x_2)\hat{\pi}_H(x) \int_{-\infty}^{+t} dx_4 J(x_4)\hat{\varphi}_H(x_4) \\
&+ \int_{t}^{+\infty} dx_1 J(x_1)\hat{\varphi}_H(x_1) \int_{t}^{+t_1} dx_2 J(x_2)\hat{\varphi}_H(x_2) \int_{t}^{+t_2} dx_3 J(x_3)\hat{\varphi}_H(x_3)\hat{\pi}_H(x)
\end{aligned} \tag{5.8}$$

Take the derivative of this equation with respect to $t$ to obtain,

$$\frac{\partial^2}{\partial t^2}\frac{\delta \hat{S}_4}{\delta J(x)} = \hat{F}_1 + \hat{F}_2 \tag{5.9}$$

where,

$$
\begin{aligned}
\hat{F}_1 = \ddot{\hat{\varphi}}_H(x)\int_{-\infty}^{+t} dx_2 J(x_2)\hat{\varphi}_H(x_2)\int_{-\infty}^{+t_2} dx_3 J(x_3)\hat{\varphi}_H(x_3)\int_{-\infty}^{+t_3} dx_4 J(x_4)\hat{\varphi}_H(x_4) \\
+\int_{t}^{+\infty} dx_1 J(x_1)\hat{\varphi}_H(x_1)\ddot{\hat{\varphi}}_H(x)\int_{-\infty}^{+t} dx_3 J(x_3)\hat{\varphi}_H(x_3)\int_{-\infty}^{+t_3} dx_4 J(x_4)\hat{\varphi}_H(x_4) \\
+\int_{t}^{+\infty} dx_1 J(x_1)\hat{\varphi}_H(x_1)\int_{t}^{+t_1} dx_2 J(x_2)\hat{\varphi}_H(x_2)\ddot{\hat{\varphi}}_H(x)\int_{-\infty}^{+t} dx_4 J(x_4)\hat{\varphi}_H(x_4) \\
+\int_{t}^{+\infty} dx_1 J(x_1)\hat{\varphi}_H(x_1)\int_{t}^{+t_1} dx_2 J(x_2)\hat{\varphi}_H(x_2)\int_{t}^{+t_2} dx_3 J(x_3)\hat{\varphi}_H(x_3)\ddot{\hat{\varphi}}_H(x)
\end{aligned}
\tag{5.10}
$$

where $\ddot{\hat{\varphi}}_H(x) \equiv \dfrac{\partial^2 \hat{\varphi}_H(x)}{\partial t^2}$ and,

$$
\begin{aligned}
\hat{F}_2 = \left[\hat{\pi}_H(\vec{x},t), \int d^3x' J(\vec{x}',t)\hat{\varphi}_H(\vec{x}',t)\right]\int_{-\infty}^{t} dx_3 J(x_3)\hat{\varphi}_H(x_3)\int_{-\infty}^{t_3} dx_4 J(x_4)\hat{\varphi}_H(x_4) \\
+\int_{t}^{+\infty} dx_1 J(x_1)\hat{\varphi}_H(x_1)\left[\hat{\pi}_H(\vec{x},t), \int d^3x' J(\vec{x}',t)\hat{\varphi}_H(\vec{x}',t)\right]\int_{-\infty}^{t} dx_4 J(x_4)\hat{\varphi}_H(x_4) \\
+\int_{t}^{+\infty} dx_1 J(x_1)\hat{\varphi}_H(x_1)\int_{t}^{t_1} dx_2 J(x_2)\hat{\varphi}_H(x_2)\left[\hat{\pi}_H(\vec{x},t), \int d^3x' J(\vec{x}',t)\hat{\varphi}_H(\vec{x}',t)\right]
\end{aligned}
\tag{5.11}
$$

Rearrange terms and relabel some of the dummy variables to obtain ,

$$
\hat{F}_2 = \left[\hat{\pi}_H(\vec{x},t), \int d^3x' J(\vec{x}',t)\hat{\varphi}_H(\vec{x}',t)\right]\frac{1}{2!}T\left(\int_{-\infty}^{+\infty} dx_1 \int_{-\infty}^{+\infty} dx_2 J(x_1)J(x_2)\hat{\varphi}_H(x_1)\hat{\varphi}_H(x_2)\right)
\tag{5.12}
$$

Use (2.3) to obtain,

$$
\hat{F}_2 = -iJ(x)\frac{1}{2!}T\left(\int_{-\infty}^{+\infty} dx_1 \int_{-\infty}^{+\infty} dx_2 J(x_1)J(x_2)\hat{\varphi}_H(x_1)\hat{\varphi}_H(x_2)\right)
\tag{5.13}
$$

Also, from (5.10) we obtain,

$$
\hat{F}_1 = \frac{1}{3!}T\left(\frac{\partial^2 \hat{\varphi}_H(x)}{\partial t^2}\int_{-\infty}^{+\infty} dx_1 \int_{-\infty}^{+\infty} dx_2 \int_{-\infty}^{+\infty} dx_3 J(x_1)\hat{\varphi}_H(x_1)J(x_2)\hat{\varphi}_H(x_2)J(x_3)\hat{\varphi}_H(x_3)\right)
\tag{5.14}
$$

These results are consistent with (5.5).